\documentclass[12pt]{article}

\usepackage{fullpage}
\usepackage{graphicx}
\usepackage[T1]{fontenc} 
\usepackage{xcolor}
\usepackage{ulem}

\title{Quantitative near-field characterization of Surface Plasmon Polaritons on nanofabricated transmission structure}

\begin{document}

\author{Aleksandra Sierant\textsuperscript{1{*}}, Benedykt R. Jany\textsuperscript{1}, Tomasz Kawalec\textsuperscript{1}}

\maketitle

1. Institute of Physics, Jagiellonian University in Krak\'ow, \L{}ojasiewicza 11, 30-348 Krak\'ow, Poland \\
 {*}aleksandra.plawecka@doctoral.uj.edu.pl

\maketitle

\begin{abstract}
The tailoring of plasmonic near-fields is central to the field of nanophotonics.
The detailed knowledge of the field distribution is crucial for a design and fabrication of plasmonic sensors, detectors, photovoltaics, plasmon-based cicuits, nanomanipulators, electrooptic plasmonic modulators and atomic devices. We report on a fully quantitative comparison between near field observation and numerical calculations, considering the intensity distribution for TM and TE polarisations, necessary for the construction of devices in all these areas. We present the near field scanning microscopy (NSOM) results of Surface Plasmon Polaritons (SPPs), excited by linearly polarized illumination on a gold, nanofabricated transmission grating. The optimization process is performed for infrared light, for future applications in cold atoms trapping and plasmonic sensing. 
We show the \textit{in situ} processes of build up and propagation of SPPs and
confirm that the out of plane component is not coupled to the aperture-type NSOM probe.
\end{abstract}

Near-field scanning microscopy (NSOM) is a well known imaging technique for optical near-field examination. Nowadays, it is one of
the most powerful imaging tools thanks to a wide range of probe types and achievable 
subwavelength resolutions \cite{pohl1984,durig1986,betzig1987}. The technique provides a unique opportunity to investigate the physical processes such as the extraordinary optical transmission 
\cite{gay2006optical}, light propagation in photonic crystal waveguides
\cite{bozhevolnyi2002near}, and dynamics of plasmonic nanoantennas 
\cite{dorfmuller2011near,schnell2009controlling}. 
The knowledge of the distribution of near
field allows for the design of a number of devices, e.g. nanomanipulators 
\cite{minovich2011generation,juan2011plasmon}, sensors based on surface enhanced Raman spectroscopy 
\cite{li2010shell}, plasmonic sensors \cite{sensory}, surface plasmons-based circuits \cite{Ozbay189}, electro-optic plasmonic modulators \cite{cai2009}, and chip-scale atomic devices \cite{chip}. 
The understanding of near fields is necessary for the construction of structured optical potentials
for cold atoms, e.g. atom mirrors and plasmonic surface traps
\cite{stehle2011plasmonically,mildner2018plasmonic,kawalec2017,kawalec2014}. 

Surface Plasmon Polaritons (SPPs), which emerge from a coupling between the light and collective oscillations of free electrons at a metal surface, underlie a significant number of the above-mentioned experiments. The SPPs propagate along the metal-dielectric boundary, and the
amplitude of the electromagnetic field is exponentially decaying in both media. The non propagating form of SPPs, Localized Surface Plasmons, are found in the vicinity of 
metallic nanoobjects \cite{maier2007plasmonics}. Both forms of SPPs have been studied by NSOM technique \cite[and references therein]{barnes2003nature}.
The research include mapping the near field associated with a metallic waveguides based on:
stripes \cite{weeber2001near,dallapiccola2009near}, nanowires \cite{krenn2002non,wiecha2019direct},
cavities \cite{balci2011direct}, nanoparticles \cite{krenn1999squeezing, maier2003local}, or
gold film surface covered with randomly positioned scatterers
\cite{bozhevolnyi2002localization,bozhevolnyi2003local}. The plasmon modes
were also imaged in gold nanorods \cite{imura2005near}, gratings 
\cite{jose2011enhanced,iqbal2015propagation,gadalla2020imaging}, slits 
\cite{kihm2008control,venuthurumilli2019near,costantini2012situ}, metallic 
discontinuities \cite{salomon2002local}, Au-Al heterostructure \cite{kitazawa2007snom}, 
and nanoholes \cite{yin2004surface,he2011near,chang2005surface}. Apart from localized
and propagating SPPs, also the standing waves of SPPs, have been observed by NSOM
technique for a variety of structures \cite[and references therein]{ye2014optical}, 
including a set of nano-slits \cite{dvorak2013control,dvorak}.

One of the available optical methods of generating SPPs, uses a grating coupler, 
which, unlike any prism-based configuration, allows to miniaturize the system \cite{maier2007plasmonics}. 
The idea is to match the momenta of the 
incident light of TM polarization state and that of the SPPs. Among the wide selection of various
grating types, the transmission structures are of particular interest, because it is possible to generate the SPP on both, glass-gold and gold-air 
boundaries. The latter is advantageous for cold atom experiments and NSOM imaging. Atomic mirrors use SPPs' evanescent field to create a strong, repulsive potential
for an atomic cloud \cite{stehle2011plasmonically,mildner2018plasmonic,kawalec2014,kawalec2017} and the presence of light on the gold-air side leads to unwanted atom-photon scattering. Likewise, it would be detected by the probe, disturbing the NSOM observation of SPPs' themselves. 

So far, only  qualitative studies of the absolute 
enhancement of the electromagnetic field have been presented, as we believe. Here, we present a fully 
quantitative analysis of the SPPs' excitation on nanofabricated transmission structure. 
The NSOM imaging is confronted with numerical models, revealing the \textit{in situ} processes of build 
up and propagation of SPPs on a large-area grating coupler.\\

\begin{figure}
\centering
\includegraphics[width=0.8\textwidth]{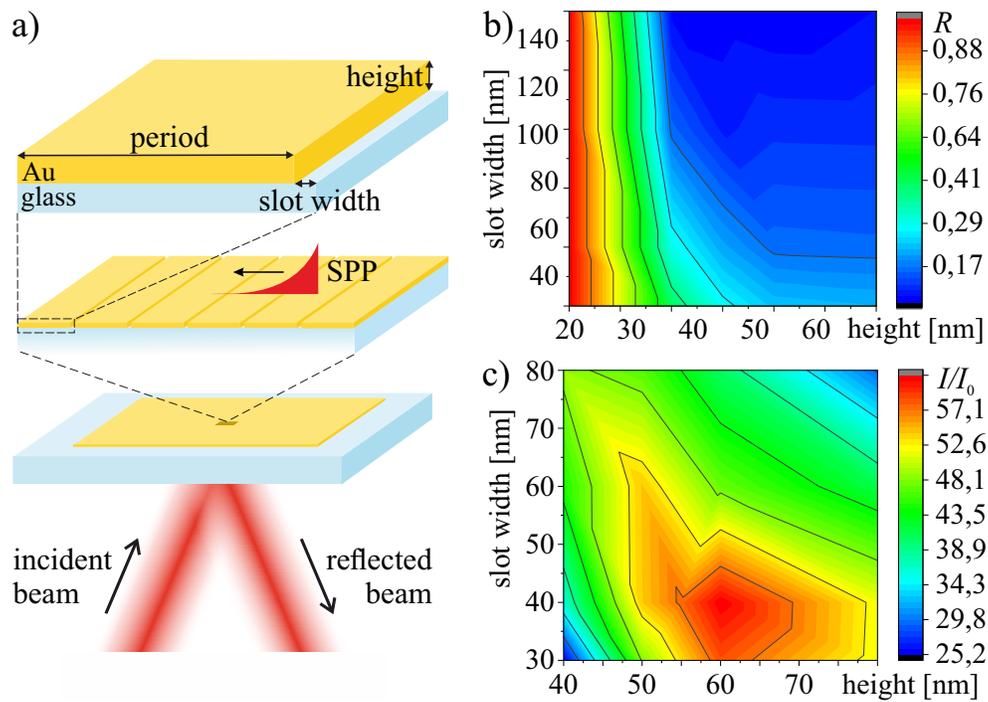}
\caption{Optimization performed by RCWA method. a) Scheme of the modeled, gold 
transmission structure, with one grating period magnified. b) Reflectivity and c) 
intensity maps for the grating of 550 nm grating period for various slot
widths and grating heights.}
\label{figure_1}      
\end{figure}

\begin{figure}
\includegraphics[width=\textwidth]{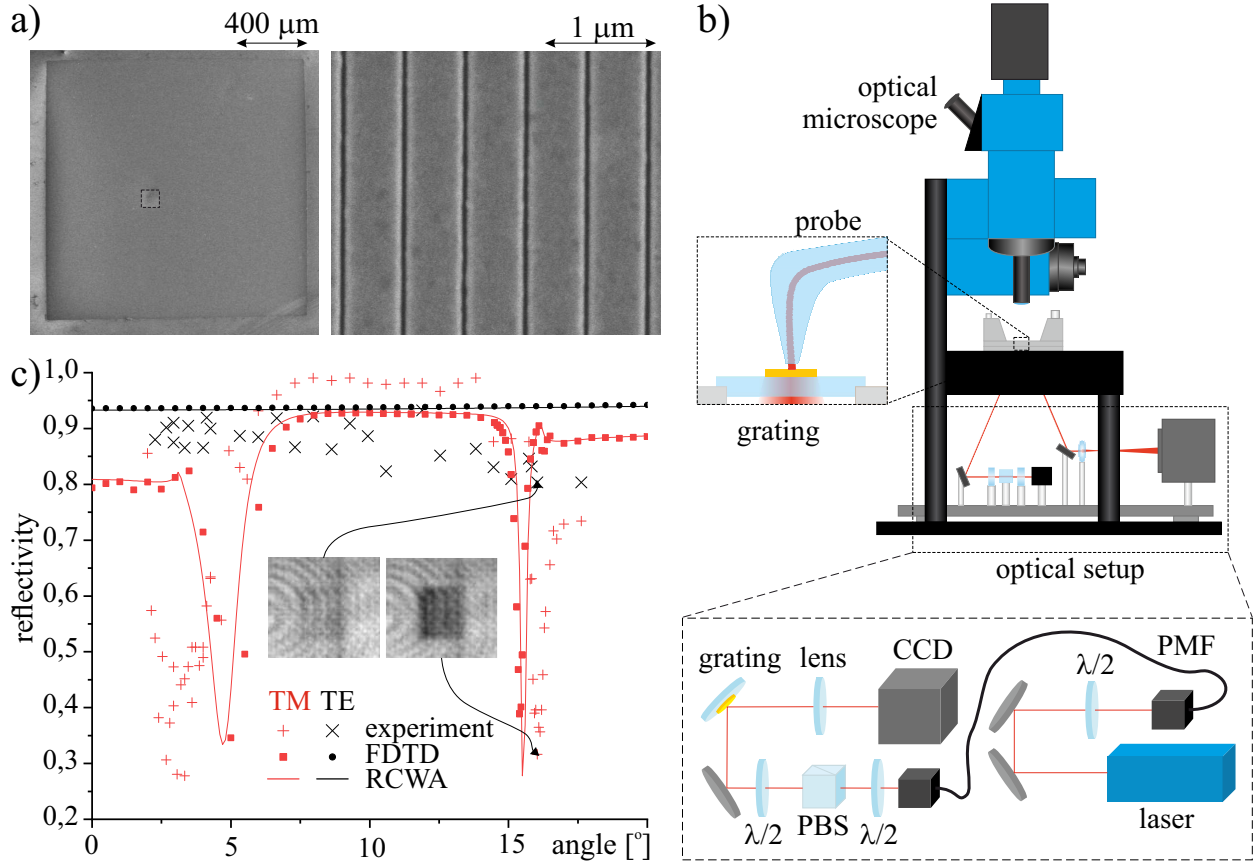}
\caption{a) SEM micrographs of the transmission diffraction grating: the position of the
grating on the Au/Glass substrate and detailed view of 5 grating periods. Grating is
marked by dashed line. b) Schematic of the NSOM installed above the breadboard with the
optical setup. $\lambda/2$ is a half waveplate, PBS is polarization beam splitter, 
PMF is polarization maintaining fiber. c) Angle dependent zeroth order reflectivity 
of the 780 nm laser beam: calculations performed by FDTD (circles and squares) and RCWA (solid lines) methods, compared with experiment (crosses and pluses) for TE (black) 
and TM (red) polarizations of light. In the inset: photos of the reflected beam for
the optimum angle of incidence for TE and TM polarized light, imaged by a CCD camera.}
\label{figure_2}       
\end{figure}

To maximize the efficiency of the excitation process, we have numerically optimized the 
parameters of the grating structure, i.e. grating period, slot width, and grating height,
to obtain a narrow and deep plasmonic resonance, with a strong electromagnetic field 
enhancement above the grating surface. We focus on narrow-width slots, as proposed by Yoon \cite{yoon2006}. The modeled grating geometry is presented in Fig. \ref{figure_1}(a).
We have performed in-depth calculations of reflectivity coefficient $R$ and 
electromagnetic field intensity $I/I_0$, where $I_0$ is the intensity of the incident light, via Rigorous Coupled Wave Analysis, see Supporting Information for more information. The most effective plasmonic resonance and electromagnetic
field enhancement are obtained for grating period equal to 550 nm with reflectivity and intensity shown in Fig. \ref{figure_1} (b) and (c).
The reflectivity coefficient $R$ is minimized
for slots width 100-140 nm, and heights between 
40-80 nm. On the other hand, taking into account the intensity of the electromagnetic 
field, we set the optimal range to 30-50 nm wide slots. Due to 
the technically demanding fabrication process, especially in the case of large area structures, the final parameters have
been tuned so that the production errors (of the order of a 
few nm) do not affect the resonance significantly. Finally, 550 nm of 
grating period, 55 nm of grating height and 40 nm of slot width have been 
chosen for fabrication, and implemented in all calculations discussed in 
this article. The structure was nanofabricated by Focused Ion Beam (FIB) milling technique,
the details are contained in Supporting Information. Directly after the FIB processing,
the grating was examined in a Scanning Electron Microscope (SEM) at 5 keV electron energy in the same apparatus.
The SEM images of the grating are presented in Fig. \ref{figure_2}(a).
The detailed SEM analysis shown that the 550 nm grating period of the structure and
38 nm wide slot were achieved. In addition, presence of 
spherical like shaped grains, with radius up
to 14 nm was revealed.\\

Measurements of the optical response of the grating were carried out using a setup shown in Fig. 
\ref{figure_2}(b), details of which are described in the Supporting Information. The SPPs were excited on the
gold-air boundary and the transversal intensity distribution of the reflected light intensity was monitored on a CCD camera to control the
plasmonic resonance efficiency. At the same time, the gold-air boundary was scanned by 
NSOM fiber tip to image the electromagnetic field intensity and surface topography. The presence of the SPPs is demonstrated through an extinction in the reflected 
light intensity, caused by a phase difference between specularly reflected and radiated 
light (see the inset in Fig. \ref{figure_2}(c)). The measured coefficient $R$, compared with 
numerical simulations, is presented in Fig. \ref{figure_2}
(c). The Rigorous Coupled Wave Analysis (RCWA) assumes a perfectly flat, rectangular grating, whereas the Finite Difference Time Domain (FDTD), takes into account the irregularities of the
gold surface. The minimum located around 4$^\circ$ results from the SPPs generated 
on the glass-gold boundary. The second one, located around 16$^\circ$, corresponds to the SPPs excited on the gold-air edge with experimentally achieved reflectivity equal to 0.32 (coupling efficiency of 68$\%$). The narrow and deep resonance makes the grating an effective scientific tool for plasmonic sensing. The change in a refractive index of $\Delta n=0.00002$, which
is the difference between the refractive indices of air and nitrogen at 780 nm, 
will result in the reflectivity decrease of 0.3$\%$, which can be easily detected
in amplitude-type sensors. A very good agreement between the simulations performed by RCWA and FDTD 
methods is observed, with a mismatch in the $R$ coefficient calculated 
at the optimal angle of incidence 
($R_{RCWA}=0.26$ and $R_{FDTD}=0.39$). 
The difference between $R_{RCWA}$ and $R_{FDTD}$ quantifies the degree to which the plasmonic resonance is 
deteriorated by the surface imperfections.
The periodic boundary 
conditions imposed in numerical calculations make the modeled grating to be virtually 
an \textit{infinite} structure, are a source of discrepancies between the simulations and experimental points. 
The real grating consisted of 100 $\mu$m of grating 
periods, surrounded by flat gold surface (hereinafter refereed to as a \textit{finite} grating).\\

The near field distributions were imaged by NSOM, details of which are described in the Supporting Information. Scanning of the entire sample's area and tens of $\mu$m beyond was performed, with a single scan size $20\times20$ $\mu$m. For each measurement we have performed FDTD simulations, taking into account the finite size of the grating: the modelled structure consisted of 183 grating periods surrounded by 100 $\mu$m of a plain gold from both sides. The surface corrugations were included in the calculations. \\

\begin{figure}
\includegraphics[width=\textwidth]{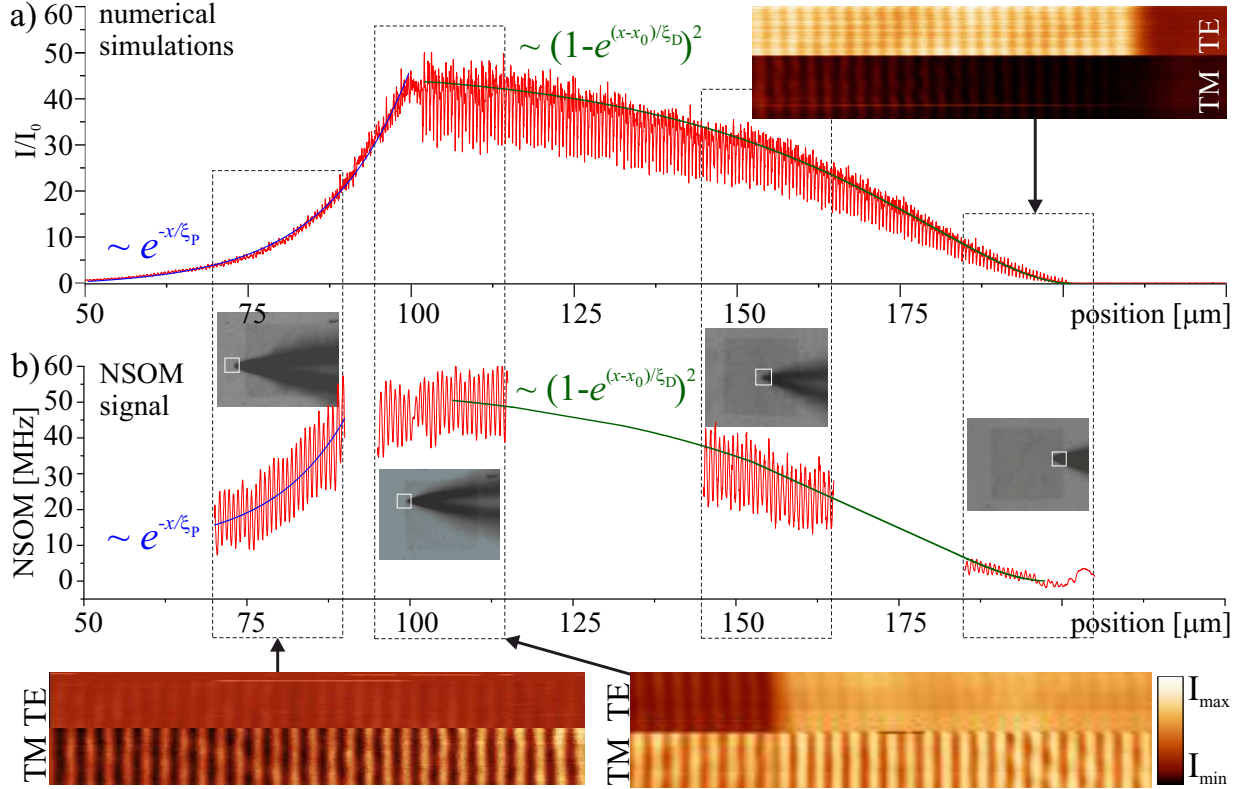}
\caption{Electromagnetic field intensity above the transmission diffraction grating  
calculated by FDTD method (a) and measured by NSOM (b). Note the 
separate intensity scales in both graphs. Measurements have been performed
on the right, left and outside the left edge of the grating, as indicated in 
the graphs by images taken by the optical microscope. The 
figure is completed with NSOM images of the respective areas.}
\label{figure_5}       
\end{figure}

The analysed areas are shown in Fig. \ref{figure_5}. We distinguish: \textit{the right edge}, \textit{the center}, \textit{the left edge}, and \textit{outside the left edge}. 
The light that excited SPPs illuminated the sample from the left side. 
Fig. \ref{figure_5} (a) presents FDTD simulations of the relative 
intensity $I/I_0$, for TM state of
polarization, compared with the enhancement of electromagnetic field
measured with NSOM exhibited in Fig. \ref{figure_5} (b). The scanned
area is shown 
by the images, taken by the optical microscope, and completed with
an appropriate NSOM image for both states of polarization, imaged together
in a single measurement cycle (in the middle of the single $20\times20$ $\mu$m scan, the light polarization was changed). The process of SPPs' build up takes place at
the right side of the grating, demonstrated by the very low intensity for TM 
polarized light and poor contrast of the fringes. Then, in the middle of 
the grating, the signal and the contrast become stronger.  
The SPPs propagate even 
outside the left side of the grating,  the enhancement of intensity
reaches a factor of 50.
Although the sample has no grooves outside the grating, 
the fringes are clearly visible -- this is due to the interference 
between the propagating SPPs, and the incident beam, which leaks 
through the gold layer. Results of the numerical simulations match
qualitatively
the intensity distribution measured by NSOM. A quantitative agreement for the central region will be shown later. The intensity profile allows to determine the propagation length $\xi_P$ of SPP
excited on the structure, which describes the distance, at which the intensity 
of SPPs decreases $e$ times \cite{raether1988surface}. The analytically calculated
propagation length, for flat, gold surface of infinite thickness is $\xi^{flat}_P=43$ $\mu$m 
\cite{raether1988surface}. In the case of thin layers, apart from the always-present 
ohmic losses, also the leakage radiation reduces the propagation length approximately by a factor of 2, what is as confirmed in our numerical simulations for a perfect grating (no irregularities
on the surface), giving $\xi^{perf}_P=21$ $\mu$m. However, to determine the propagation length in a realistic scenario, the irregularities of the grating surface
must be taken into account. 
The exponential function $\sim \exp{(-x/\xi_{P})}$ has been fitted to both, 
numerical simulations, and NSOM signal (denoted in Fig.~\ref{figure_5}
by the blue lines). 
The numerically calculated propagation length is $\xi^{FDTD}_P=12.9(0.1)$ $\mu$m and is consistent with the
experimentally measured value $\xi^{EXP}_P=11.5(2.0)$ $\mu$m. 
Furthermore, the process of SPPs' build up is investigated.
Both numerical and experimental intensity profiles
are very well described by an analytic formula \cite{koev2012efficient}
$I/I_0\sim (1-\exp{((x-x_0)/\xi_{D}))^2}$, where $\xi_{D}$
is the decay length. The calculated value of the decay length is $\xi^{EXP}_D=31(2)$
$\mu$m for NSOM signal, consistent with the value $\xi^{FDTD}_D=30.3(0.1)$ $\mu$m from numerical simulations.

\begin{figure}
\includegraphics[width=\textwidth]{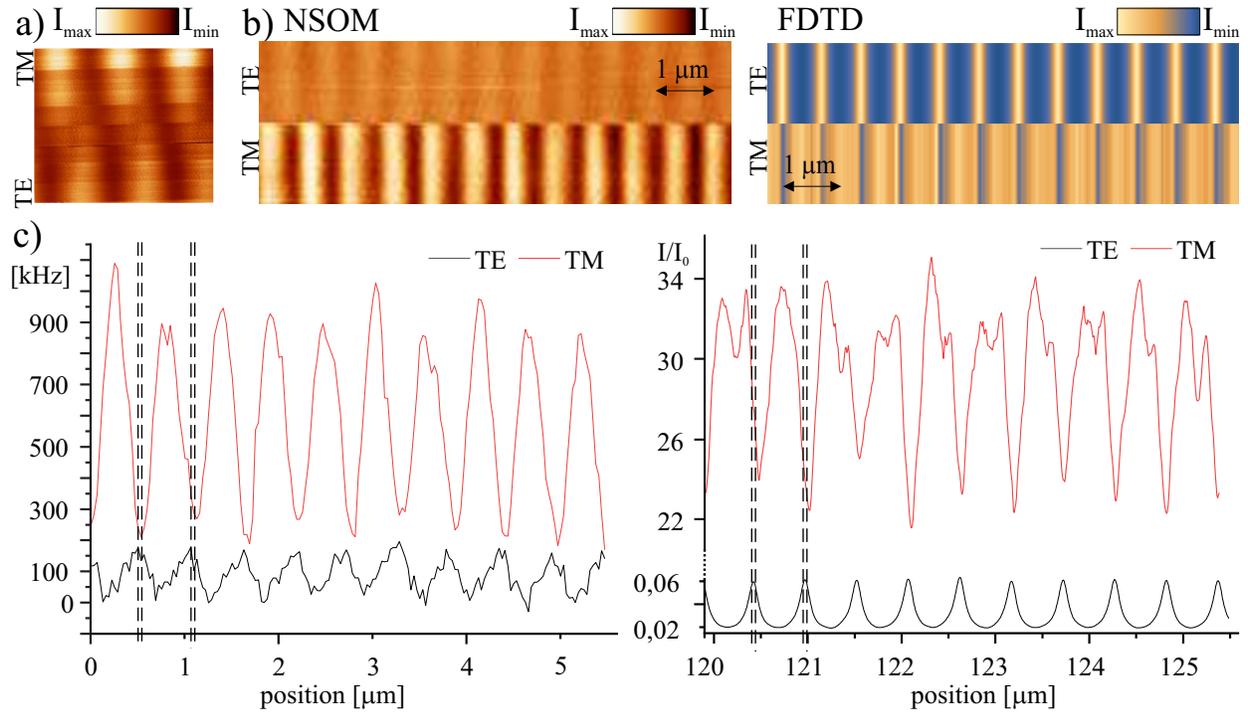}
\caption{Electromagnetic field enhancement above the transmission
diffraction grating measured by NSOM and numerically simulated by
FDTD method for TE and TM polarized light. a) Qualitative comparison 
for polarization of light varying between TE and TM, together in a single 
measurement cycle. b) Comparison between TM and TE polarizations -- 
direct NSOM result (left) and respective FDTD simulations (right). c) 
Quantitative analysis of (b): light intensity at 
NSOM and FDTD cross-sections, obtained 
for TM (red line) and TE (black line, note the double vertical scale)
polarizations of light. Two of the grating slots, based on the AFM 
topography, are denoted by black dashed lines, given as a reference.
}
\label{figure_3}      
\end{figure}

\begin{figure}
\includegraphics[width=1\textwidth]{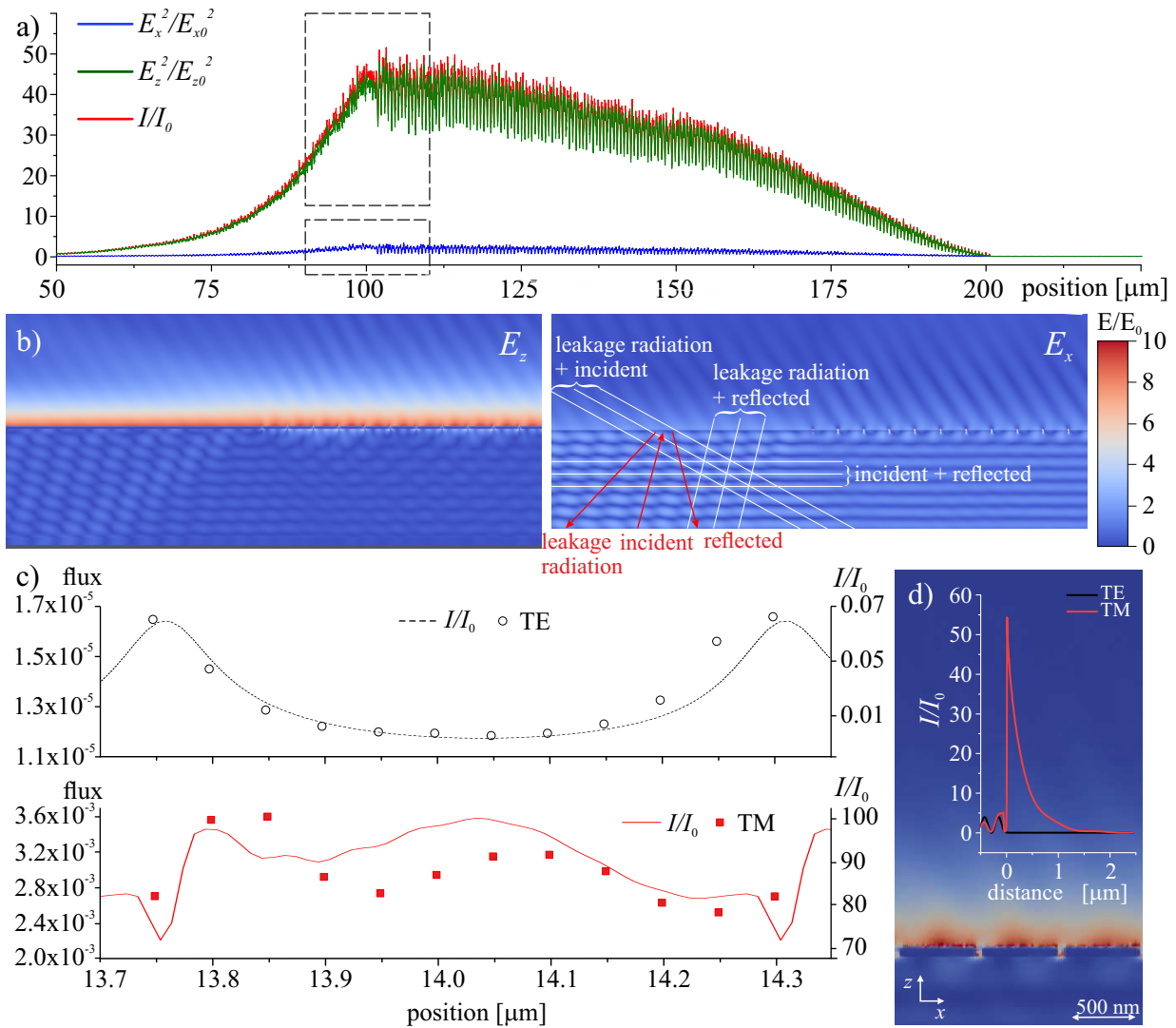}
\caption{FDTD simulations of electromagnetic field enhancement above the transmission diffraction grating.
a) The quantitative comparison between components $E_x$ (in plane), $E_z$ (out of plane) and total intensity 
$I$, for TM polarized light. b) Amplitude of the respective components on the left edge of the grating.  
The maxima of interference pattern, resulting from interference between the incident and reflected light 
and leakage radiation, are schematically marked with white lines. b) Poynting flux calculated for TM 
(red squares) and TE (black circles) polarized light, compared with the intensity distribution
calculated without the probe presence (solid lines). c) Simulated $I/I_0$ distribution for TM 
(up) and TE (down) polarized light. Inset: Cross section in $z$ direction for TE (black line) and 
TM (red line) polarized light.}
\label{figure_4}       
\end{figure}

Fig. \ref{figure_3} compares the results obtained for different states of 
polarization, varying between TM (SPP presence) and TE (SPP absence) states, 
taken at the center of the grating.
A qualitative demonstration is presented in Fig.~\ref{figure_3} (a), 
showing the signal detected by NSOM, during a change of
polarization state. The state was varying from TM to TE polarization, 
by rotating the orientation of the plane of polarization, with the step
of 7.5$^\circ$. The analysis demonstrates a significant change of the 
intensity distribution -- the maxima and minima switch their positions,
and the strength of the signal weakens with decreasing SPP excitation
(that is, from TM to TE). Results of NSOM measurements and FDTD simulations for TM and TE polarization states
are shown in Fig.~\ref{figure_3} (b), cross-sections allowing for a quantitative
comparison are exhibited in Fig.~\ref{figure_3} (c).
Note, that 
the numerical simulations have been performed for the finite grating, and 
the results are presented for only 13 grating periods, taken from the 
middle of the grating. The position of minima and
maxima with respect to grating's ridges and grooves are well matched 
for the simulations and experiment. The maxima of the TM 
case occur in the middle of the ridge, while the maxima of TE polarized 
light can be found in the vicinity of the groove. Some discrepancies can 
be found in the shape of the maxima which are slightly wider in the NSOM data. This is caused by the spatial integration of the collected signal (note, that the 
diameter of the fiber aperture is 50 nm, excluding the Cr-Au cladding, and 
the width of the slot is 40 nm). Furthermore,  the TM case gives stronger signal than TE, leading to  a typical TM/TE
ratio measured by NSOM to be 10, and the highest recorded ratio was 40. 
According to the FDTD simulations, 
the TM/TE ratio is expected to be much higher, around 500.
This discrepancy, in favour of the TE polarization intensity can be readily understood.
According to the Rotenberg \textit{at al}, the \textit{out of plane} 
electric field component is greatly suppressed in NSOM detection, as opposed 
to the \textit{in plane} component, for aperture-type probes \cite{rotenberg2014mapping}.
The electric field emerging from the TM polarized light has two components: $E_x$ 
and $E_z$, which are the in plane, and out of plane components,
respectively. The 
electric field arising due to the TE polarized light has only $E_y$ component,
which is the in plane component detected efficiently by the 
NSOM probe. It already explains qualitatively the nature of the coupling of both polarizations to the
fiber tip. A quantitative approach is shown in Fig. \ref{figure_4} (a) and (b).
The main plasmonic enhancement is due to the $E_z$
component, which explains the differences 
between the  TM/TE ratios obtained in the simulations and NSOM measurements. 
Results in Fig.~\ref{figure_4} (a) show that the $E_z^2$ 
component (green line) is 20 times stronger 
than the $E_x^2$ component (blue line), so that $E_z^2$ almost completely overlaps with the total intensity distribution (red line). This is an inherent property of the SPPs
on gold for the considered wavelength of 780 nm, 
in contrast to the case of the pure evanescent wave, where 
the contribution of the components may be easily varied \cite{jozefowski2007direct}.
The strong domination of the $E_x$ component in the NSOM results explains the well pronounced interference pattern measured outside the grating, 
not predicted in the simulations 
(compare Fig. 3 (a) and (b), left side of the grating).

The interference pattern in the glass substrate volume has an interesting origin. 
It comes from the interference between three waves: the incident beam, the beam 
reflected from the gold surface and the leakage radiation, as depicted in Fig. \ref{figure_4} (b) 
(c.f. \cite{fiutowski2014leakage}). The angle of incidence of the laser light is 15.4$^\circ$ 
and the angle of the leakage radiation is 43$^\circ$ in glass.
The intensity distribution above 3 grating periods and the respective cross section 
are presented in Fig.~\ref{figure_4} (d). 
As expected, the exponential decay 
from the surface is revealed, thus proving the usefulness of such a structure in cold atoms experiments.
In order to explain the discrepancy between FDTD intensity distribution and NSOM results,
we have taken into account the presence of the probe. Motivated by Dvo\v{r}ak \textit{et al.}, 
we have performed the Poynting flux calculations inside the fiber probe, taking into
consideration the detection of individual field components \cite{dvorak2013control}. 
The calculation model consisted of 20 $\mu$m long row of grating periods, with the 
NSOM probe above them (note, that this is the \textit{infinite} type of the structure;
the calculations for the finite case were infeasible.
The NSOM probe consisted of a glass structure of a 50 nm diameter, covered with 100 nm of
gold coating and was placed 15 nm above the grating structure. After calculating the flux in one
position, the simulations have been repeated for the probe shifted horizontally in 50 nm 
increments, scanning more than one grating period in total. The summary is presented in Fig. 
\ref{figure_4} (c), by comparing the value of Poynting flux and the intensity distribution 
$I/I_0$ (which was calculated without the NSOM probe, but for the very same 20 $\mu$m of 
grating periods, also for the \textit{infinite} type of the grating). The 
shape of the intensity distribution is indeed reproduced
for both polarization states, thus, it matches the shape of the NSOM signal. The TM to TE ratio 
determined by $I/I_0$ equals 2000, while the one driven by Poynting flux gives factor 
of 200, reducing the no-probe ratio ten times, giving the previously missing factor. 
This proves, that the out of plane component is not detected by NSOM, and shows that 
our quantitative comparison between NSOM measurement and FDTD simulations is very accurate.\\

In summary, we have used a NSOM technique to investigate the nature of SPPs on a 
nanofabricated transmission grating. We have optimized the parameters of the grating
and we have fabricated the structure with FIB microscope. Our far field 
goniometric measurements revealed the narrow and deep plasmonic resonance, with 
an efficiency coupling of 68$\%$, and electromagnetic field enhancement of 50 
This makes the grating an effective scientific tool for plasmonic sensing and a promising base for cold atom experiments, including precise micropotentials tailored with subwavelength resolution through SPPs excitation. The proposed transmission structure allows for the miniaturization of the system, ensures the separation between the exciting beam from the proper part of the experiment, thus can be successfully used in optical dipole mirrors, surface traps, and other plasmonic devices
\cite{stehle2011plasmonically,mildner2018plasmonic}. We have 
directly observed the \textit{in situ} processes of SPPs building up with the SPP decay length of  $\xi^{EXP}_D=31$ $\mu$m and the propagation outside the grating structure, with the propagation length equal to $\xi^{EXP}_P=11.5$ $\mu$m. We have confirmed, that the aperture-type probe does not couple the out of plane component of the electric field. We have also quantitatively compared the signal detected by NSOM and FDTD calculation, reaching the enhancement of 50 between the excited electromagnetic field and the incident light. We have measured the intensity distribution associated with propagation of SPPs and demonstrated an accurate agreement of numerical simulations with experimental data. Our simulations also show that it is crucial to take into account the microscopic details of the structure such as the irregularities of the gold surface and the finite size of the grating for quantitative understanding of propagation and decay of SPPs. This paves the way for the full quantitative understanding of SPPs needed for a design and fabrication of plasmonic sensors, plasmon-based cicuits, photovoltaics, nanomanipulators, electrooptic plasmonic modulators and atomic devices.\\

We acknowledge the support of National Science Centre (doctoral scholarship – A.S.) under project 2018/28/T/ST2/00275 .\\

The authors thank Prof. Jakub Rysz, for lending the NSOM, his expert assistance, and useful discussions, and Prof. Franciszek Krok for the opportunity to use the SEM/FIB microscope. We also thank Dr. Jacek Fiutowski for preparing gold substrates, Dr. Pawe\l{} D\k{a}bczy\'nski for plasma cleaning of the grating, Dr. Dominik Wrana for SEM imaging, Dr. Dobros\l{}awa Bartoszek-Bober for contributions to the early stage of the experiment, and Dr. Piotr Sierant for remarks on the manuscript.

\section*{Supporting Information}
\subsubsection*{Structure fabrication}
The 55 nm thick gold layer was evaporated by electron beam metal deposition onto a glass substrate, supported by 3 nm titanium adhesion layer. The structure was fabricated by FIB milling (Dual Beam SEM/FIB Quanta 3D FEG microscope by FEI), using gallium ions of 30 keV energy, into a gold layer. The grooves are 38 nm wide and 100 $\mu$m long, continuously repeated by 550 nm, on the area of 100 $\mu$m $\times$ 100 $\mu$m. To prevent sample charging during FIB nanopatternig, electron flood gun charge neutraliser was used. 

\subsubsection*{Optimization of structure parameters} 
The calculations for the optimization shown in Fig. \ref{figure_1} (b) and (c) were calculated by Rigorous Coupled Wave Analysis (RCWA), using \textit{rcwa-1d}, by Pavel Kwiecien from Czech Technical University in Prague. The calculations were performed in the regime of monochromatic, near-infrared 780 nm laser light, with complex refractive index of gold equal 0.1478-4.6223$i$, and the refractive index of glass equal 1.51. In order to avoid light diffraction into orders other than zeroth, the scanned range of the grating periods was 400-760 nm, slot widths of 30-140 nm, and grating heights of 20-80 nm. The periodic boundary conditions have been implemented in the calculations, making the grating to be virtually an infinite structure. 

\subsubsection*{Numerical calculations} 
The near field distributions were calculated by Finite Difference Time Domain (FDTD), using commercially available EM Explorer.

\subsubsection*{Far field measurement setup}
Details on the experimental setup are as follows. The 780 nm laser beam (Toptica DLX100) was guided through single mode,
polarization maintaining fiber, followed by the polarizer,
and reflected from the gold
transmission grating from the glass-gold boundary. The
presence of the plasmonic resonance was determined with CCD images, taken for TE and TM polarized laser 
beam. The reflectivity coefficient, $R$, was calculated as a ratio of the intensity of the light at the grating to the intensity of the light reflected from the plain gold. The measurements were repeated for
each angle separately, up to 20$^\circ$. 

\subsubsection*{Near field measurement setup}
The intensity of the electromagnetic field in the vicinity of the grating was measured with Nanonics MultiView 1000 Scanning Near Field Microscope. The examined area was 
observed in real time through an optical microscope integrated with the NSOM system. The NSOM has been working in contact, in a collection mode, with multimode Cr-Au coated fibers of two diameters: 50 and 100 nm. The collected data was analysed and compared with the AFM topography registered simultaneously by the probe.
 
\bibliographystyle{unsrt}
\bibliography{Sierant_manuscript}

\end{document}